\documentclass[a4paper]{jpconf}
\usepackage{latexsym}
\usepackage{amsmath}
\usepackage{amssymb}
\usepackage{amsfonts}

\usepackage[mathscr,scaled=1.15]{urwchancal}
\DeclareFontFamily{OT1}{pzc}{}
\DeclareFontShape{OT1}{pzc}{m}{it}%
{<-> s * [1.15] pzcmi7t}{}
\DeclareMathAlphabet{\mathpzc}{OT1}{pzc}{m}{it}

\usepackage{color}

\usepackage{supertabular}
\usepackage{placeins}
\usepackage{epsfig}
\usepackage{graphicx}
\usepackage{cite}

\definecolor{prdblue}{rgb}{0.133,0.118,0.498}
\usepackage[colorlinks=true, pdfstartview=FitV, linkcolor=prdblue, citecolor= prdblue, urlcolor=prdblue]{hyperref}

\def\lsim{\raise0.3ex\hbox{$<$\kern-0.75em\raise-1.1ex\hbox{$\sim$}}}
\def\gsim{\raise0.3ex\hbox{$>$\kern-0.75em\raise-1.1ex\hbox{$\sim$}}}

\begin{document}

\begin{flushright}
NJU-INP 006/19
\end{flushright}

\title{Process-independent effective coupling and the pion structure function}

\author{Jos\'e Rodr\'{\i}guez-Quintero$^{1}$, Lei Chang$^{2}$, Kh\'epani Raya$^{2}$ \\ and Craig D. Roberts$^{3,4}$}
\address{$^{1}$ Dept. of Integrated Sciences and Center for Advanced Studies in Physics, Mathematics and computation, University of Huelva, E-21071 Huelva, Spain}
\address{$^{2}$ School of Physics, Nankai University, Tianjin 300071, China}
\address{$^{3}$ School of Physics, Nanjing University, Nanjing, Jiangsu 210093, China}
\address{$^{4}$ Institute for Nonperturbative Physics, Nanjing University, Nanjing, Jiangsu 210093, China}
%\ead{leichang@nankai.edu.cn}
%\ead{khepani@gmail.com}
%\ead{cdroberts.inp@gmail.com}
\ead{jose.rodriguez@dfaie.uhu.es}

\begin{abstract}
We sketch the calculation of the pion structure functions within the DSE framework, following two alternative albeit consistent approaches, and   . discuss then their QCD evolution, the running driven by an effective charge, from a hadronic scale up to any larger one accessible to experiment. 
\end{abstract}

\section{Introduction}

Pions, the Nature's simplest hadrons, are simultaneously Nambu-Goldstone modes generated by dynamical chiral symmetry breaking in the Standard Model (SM) and bound states of first-generation light quarks and anti-quarks. This key feature explains why symmetries and their breaking play a crucial role in accounting for pions' properties. More importantly, it is also why charting and understanding pions' structure and mass distribution in terms of SM strong interactions is a cumbersome, central problem in modern physics, demanding a coherent effort both in QCD continuum and lattice calculations and in experiments shedding a light on this understanding \cite{Aguilar:2019teb}. 

A basic quantity revealing the pion's structure is its parton distribution function, ${\mathpzc q}^{\pi}(x;\zeta)$, expressing the probability that a ${\mathpzc q}$-flavour valence quark carries a light-front momentum fraction $x$ in the pion. In particular, this density has been the object of a long controversy, since that leading-order perturbative QCD analysis of $\pi N$ Drell-Yan data (E615 experiment \cite{Conway:1989fs}) drew as a conclusion that, at the relevant energy scale for the experiment, $\zeta_5$=5.2 GeV, ${\mathpzc q}^\pi(x;\zeta_5) \sim (1-x)$ when $x \to 1$; in clear contradiction with the result early predicted from parton model and perturbative QCD \cite{Ezawa:1974wm, Farrar:1975yb, Berger:1979du}: ${\mathpzc q}^\pi(x;\zeta_H) \sim (1-x)^2$, where $\zeta_H$ is an energy scale characteristic of nonperturbative dynamics; while QCD evolution is expected to make the exponent increase by the effect of the logarithmic running and thus become effectively $2+\gamma$, with $\gamma \gtrsim 0$, for any scale $\zeta > \zeta_H$. Subsequent continuum QCD calculations~\cite{Hecht:2000xa,Chang:2014lva,Ding:2019lwe} and further careful re-analyses of E615 data~\cite{Wijesooriya:2005ir, Aicher:2010cb}, including soft-gluon resummation, have recently reported results consistent with an exponent equal to 2+$\gamma$; while other calculations disregarding symmetry-preserving diagrams~\cite{Bednar:2018mtf} or data analysis not yet including relevant threshold resummation effects~\cite{Barry:2018ort} claimed to have found an exponent closely around 1. 

As discussed in Ref.~\cite{Ding:2019lwe}, two key issues in determining the pion's parton distribution function, ${\mathpzc q}^\pi(x;\zeta)$, are: (i) accounting, beyond the impulse approximation, for a class of corrections to the handbag-diagram representation of the virtual-photon-pion forward Compton scattering amplitude, restoring basic symmetries in the calculation of parton distributions~\cite{Chang:2014lva,Mezrag:2014jka}; and (ii) dealing adequately with the QCD evolution of these parton distributions, from the nonperturbative scale at which they have been obtained up to one accessible to experiment. In the following, we will sketch about (i) and elaborate further on the issue (ii), particularly in connection with a recently proposed process-independent effective charge ~\cite{Binosi:2016nme,Rodriguez-Quintero:2018wma}.

\section{The pion parton distribution function}

The pion's parton distribution function can be obtained on the ground of the knowledge of the dressed light-quark propagator and pion Bethe-Salpeter amplitude (BSA), computed by solving the appropriate Dyson-Schwinger and Bethe-Salpeter equations (BSE). In order to keep a natural connection for the renormalisation scale and the reference one for QCD evolution, the Dyson-Schwinger equations (DSE) should be renormalised at a typical hadronic scale, $\zeta_H$, where the dressed quasiparticles become the correct degrees-of-freedom~\cite{Gao:2017mmp,Ding:2018xwy}. Within this DSE and BSE approach but employing algebraic {\it ans\"atze}, a first study in Ref.~\cite{Chang:2014lva} yielded some new insight to the calculation by identifying the above-mentioned symmetry-preserving corrections, eventually leading to  
%---
\begin{equation}
{\mathpzc q}^\pi(x;\zeta_H)  = N_c {\rm tr}\! \int_{dk}\! \delta_n^{x}(k_\eta) 
\; n\cdot\partial_{k_\eta} \left[ \Gamma_\pi(k_\eta,-P) S(k_\eta) \right]
\Gamma_\pi(k_{\bar\eta},P)\, S(k_{\bar\eta})\,,
\label{qFULL}
\end{equation}
after implementing the appropriate truncation; where $\int_{dk}:=\int \frac{d^4}{(2\pi)^4}$ is a Poincar\'e-invariant regularisation of the integral, $\delta_n^{x}(k_\eta):= \delta(n\cdot k_\eta - x n\cdot P)$; $n$ is a light-like four-vector, $n^2=0$, $n\cdot P = -m_\pi$; and $k_\eta = k + \eta P$, $k_{\bar\eta} = k - (1-\eta) P$, $\eta\in [0,1]$; $\Gamma_\pi$ is the pion BSA, $S(k)$ is the dressed light-quark propagator, the trace is taken over spinor indices with $N_c$=3, such that, if the BSA is canonically normalised, then $\int_0^1 dx {\mathpzc q}^\pi(x;\zeta_H)$=1. Owing to Poincar\'e covariance, no observable can be expected to depend on $\eta$, \emph{i.e}.\ the definition of the relative momentum, and this can be algebraically proved from~Eq.~\eqref{qFULL}. Another important property of~Eq.~\eqref{qFULL}, that can be made apparent after straightforward algebra, is:  
${\mathpzc q}^\pi(x;\zeta_H)={\mathpzc q}^\pi(1-x;\zeta_H)$; which is the consequence of the bound system being described in terms of two identical dressed quasiparticles, in the isospin-symmetric limit.

Then, in a further recent work~\cite{Ding:2018xwy}, realistic numerical solutions of both DSE and BSE have been applied to compute the first six Mellin moments of the valence-quark parton distribution, derived from Eq.~\eqref{qFULL} as follows
%---
\begin{equation}
\langle   x^m  \rangle^\pi_{\zeta_H}   = \int_0^1dx\, x^m {\mathpzc q}^\pi(x;\zeta_H) 
 = \frac{N_c}{n\cdot P} {\rm tr}\! \int_{dk}\! \left[\frac{n\cdot k_\eta}{n\cdot P}\right]^m \Gamma_\pi(k_{\bar\eta},P)\, S(k_{\bar\eta})
 n\cdot\partial_{k_\eta} \left[ \Gamma_\pi(k_\eta,-P) S(k_\eta) \right] \; ; \label{MellinMoments}\\
\end{equation}
%
%\begin{subequations}
%\label{MellinMoments}
%\begin{align}
% \langle   x^m  \rangle^\pi_{\zeta_H}  & = \int_0^1dx\, x^m {\mathpzc q}^\pi(x;\zeta_H) \label{MellinA}\\
%%
%& = \frac{N_c}{n\cdot P} {\rm tr}\! \int_{dk}\! \left[\frac{n\cdot k_\eta}{n\cdot P}\right]^m \Gamma_\pi(k_{\bar\eta},P)\, S(k_{\bar\eta})
% n\cdot\partial_{k_\eta} \left[ \Gamma_\pi(k_\eta,-P) S(k_\eta) \right] \; ;
%\end{align}
%\end{subequations}
%---
the Schlessinger point method (SPM) has been then used to extend this set of moments and thus get a reliable approximant for any moment; and, finally, the SPM-approximant has been applied for the reconstruction of the valence-quark distribution, ${\mathpzc q}^\pi(x;\zeta_H)$\,\cite{Ding:2018xwy}. The parton distribution is therefore fully determined, within this approach, by the kernel interaction specified for both the quark-gap and Bethe-Salpeter equations. 

\begin{figure}[t!]
\begin{minipage}{18pc}
\includegraphics[width=18pc]{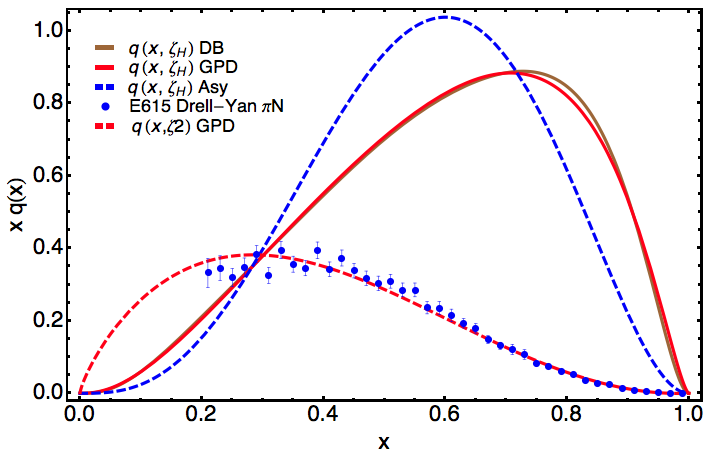}
\vspace*{-0.75cm}
\caption{\label{fig:PDFs}{\small Predicted parton distribution function from Eq.~\eqref{MellinMoments} (brown)\,\cite{Ding:2019lwe} and from Eq.~\eqref{eq:qfromLFWF} here (red) and in Ref.~\cite{Chouika:2017rzs} (blue dashed) with an algebraic model, both at the hadronic scale $\zeta_H$; evolved then up to $\zeta_5$ (red dashed), as explained in the text, and successfully compared to reanalysed E615 data\,\cite{Wijesooriya:2005ir,Aicher:2010cb} (blue circles).}}
\end{minipage}\hspace{2pc}%
\begin{minipage}{16pc}
\includegraphics[width=16pc]{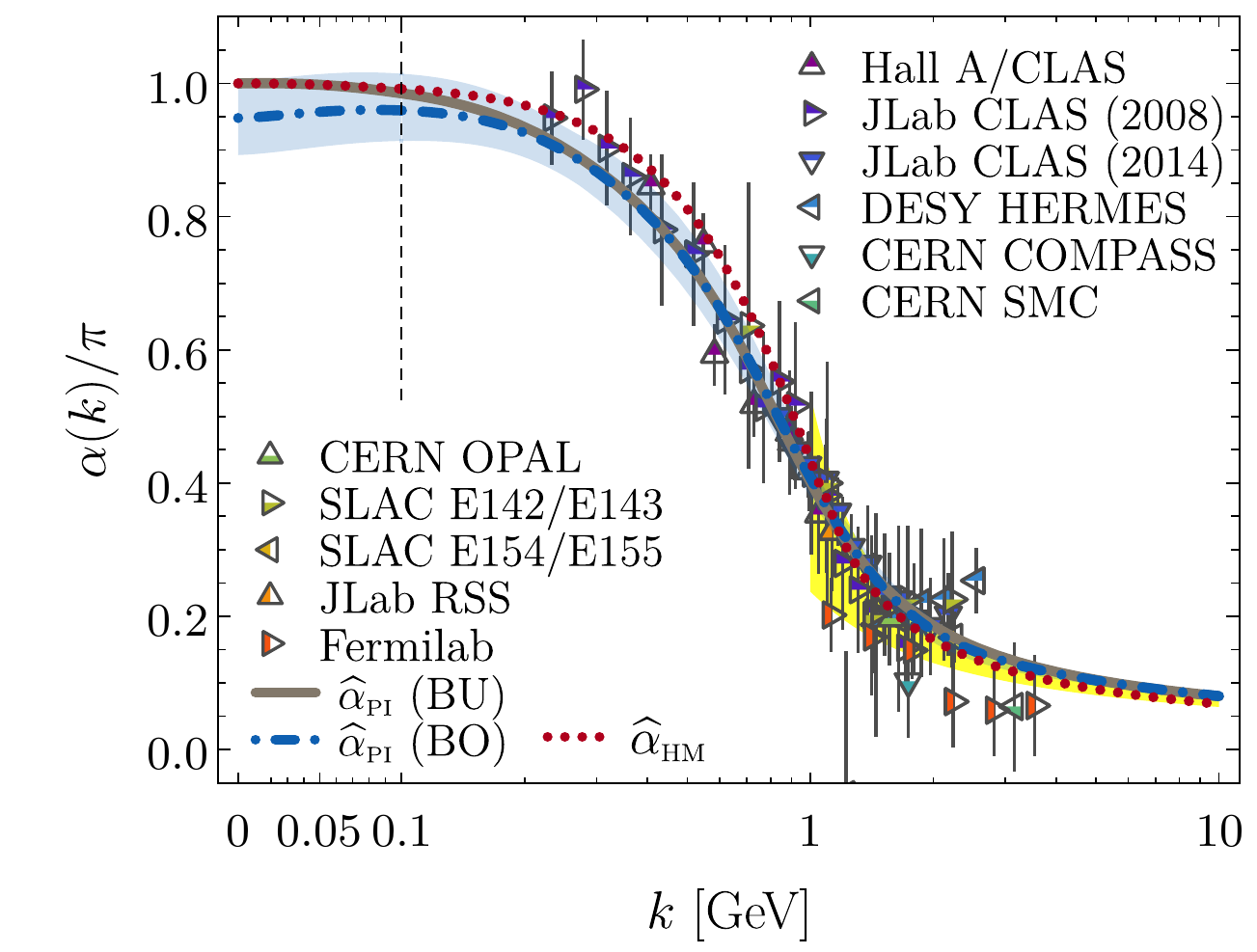}
\vspace*{-0.75cm}
\caption{\label{fig:aPI}{\small Predicted PI effective charge as obtained in Ref\,\cite{Binosi:2016nme} (dot-dashed blue) and improved in Ref.\,\cite{Rodriguez-Quintero:2018wma}, compared to the world's data of the Bjorken sum-rule charge and to the light-front holographic model (red dotted) canvassed in Ref.~\cite{Deur:2016tte}.}}
\end{minipage} 
\end{figure}

An alternative approach results from the so-called overlap representation, in which the forward limit of the generalised parton distribution gives\,\cite{Burkardt:2002uc,Diehl:2003ny}
%---
\begin{equation}\label{eq:qfromLFWF}
{\mathpzc q}^\pi(x;\zeta_H) = \int \frac{d^2 {\bf k}_\perp}{16\pi^3} \;  | \psi(x,{\bf k}_\perp^2;\zeta_H) |^2
\end{equation}
%---
for the valence-quark parton distribution in terms of the lowest Fock-space light-front wave function (LFWF) at the hadronic scale, $\psi(x,{\bf k}_\perp^2;\zeta_H)$; its leading-twist contribution resulting from the Bethe-Salpeter wave function, $\chi_\mu(k+q,k)=S(k+q)\Gamma_\pi (k+q,k) S(k)$, as
%--
\begin{equation}\label{eq:LFWF}
f_\pi \psi(x,{\bf k}^2_\perp) = \mbox{\rm tr}_{\rm CD} \int \frac{d{\bf k}_\parallel}{\pi} \; \delta_n^{x}(k) \gamma_5 \gamma \cdot n \; \chi_\mu(k-\frac{P}{2},P)\; ,
\end{equation}
%--
where $f_\pi$ is the pion's leptonic decay constant and the trace is here applied over color and spinor indices. As both the quark propagator and the BSA are in hand, basic ingredients for the realistic computation made in Ref.~\cite{Ding:2018xwy}, Eqs.~(\ref{eq:qfromLFWF},\ref{eq:LFWF}) can be also implemented to get a realistic estimate for the parton distribution within the DSE approach. Alternatively, one can follow the approach of Ref.~\cite{Xu:2018eii} and use an appropriate Nakanishi representation of the BSA, such that the LFWF eventually results from a closed expression only involving compact integrals of the so-called Nakanishi weight, a distribution defined on a support $[-1,1]$. Then, this distribution can be adjusted to reproduce the same SPM-approximant Mellin moments of Ref.~\cite{Ding:2018xwy} and, as can be seen in Fig.~\ref{fig:PDFs}, almost pointwise identical parton distributions at the hadronic scale result from both approaches. One is thus left with a realistic estimate of the LFWF which can be subsequently applied to computing the generalised parton distribution function~\cite{Rayaetal}. 

\section{DGLAP evolution of hadron structure functions}

Once the parton distribution obtained at the hadronic scale, $q^\pi(x;\zeta_H)$, one should employ the QCD-evolution equations to make it evolve up to the relevant scale for E615 and thus obtain $q^\pi(x;\zeta_5)$. The equations describing the scale violations of the hadron structure functions read
%---
\begin{eqnarray}
\left\{ 
\zeta^2 \frac{d}{d\zeta^2} 
\ - \ 
\frac{\alpha(\zeta^2)}{4\pi}
\int_x^1 \; \frac{dy}{y} \left(
\begin{array}{ccc} \displaystyle
 P_{qq}^{NS}\left(\frac x y\right) & 0 & 0 \\
0 & P_{qq}^{S}\left(\frac x y\right)  &  \displaystyle 2 n_f P_{qG}^{S}\left(\frac x y\right) \\
 0 & \displaystyle P_{Gq}^{S}\left(\frac x y\right) &  \displaystyle P_{GG}^{S} \left(\frac x y\right)
\end{array} \right) 
\right\} \;
\left( \begin{array}{c}
{\mathpzc q}^{NS}(x;\zeta) \\
q^{S}(y;\zeta) \\ 
G^{S}(y;\zeta) 
\end{array} \right) 
\ = \ 0 \ , 
\label{eq:master}
\end{eqnarray} 
%-----
written in terms of integral equations, where ${\mathpzc q}^{NS}$ stands for the non-singlet pure valence-quark and $q^S=\sum_{\mathpzc q} {\mathpzc q}+{\overline {\mathpzc q}}$ and $G^S$ represent, respectively, the singlet quark and gluon distribution functions in the pion; the elements of the matrix correspond to the so-called splitting functions as can be found, at the leading order, in Ref.~\cite{Altarelli:1977zs}, and $\alpha(\zeta)$ is the strong running coupling. Then, if the $m$-th order Mellin moment is considered, one is left with
%---
\begin{eqnarray}
\left\{
\zeta^2 \frac{d}{d\zeta^2} 
+ \frac{\alpha(\zeta^2)}{4\pi}
\left(  
\begin{array}{ccc}
 \gamma_{0,qq}^{NS,(m)} & 0 & 0 \\ 
0 & \gamma_{0,qq}^{S,(m)} & 2 n_f \gamma_{qG}^{S,(m)} \\
0 & \gamma_{0,Gq}^{S,(m)} & \gamma_{0,GG}^{S,(m)} 
\end{array} \right) 
\right\}
\left( \begin{array}{c}
\langle x_q^m \rangle_{\zeta}^{NS} \\ 
\langle x_q^m \rangle_{\zeta}^{S} \\
\langle x_G^m \rangle_{\zeta}^{S}
\end{array} \right) 
\ = \ 
0 \ ;  
\label{eq:anomalous}
\end{eqnarray} 
%----
where the coefficients for the anomalous dimension of the Mellin moments, as defined in \eqref{MellinMoments}, result from
%----
%\begin{equation}
%\label{eq:convention}
%\gamma_{0,AB}^{C,(m)} \ = \ - \; \int_0^1 \; dx \; x^m P_{0,AB}^C(x) \ ,
%\end{equation}
%----
$\gamma_{0,ij}^{k,(m)} =  - \; \int_0^1 \; dx \; x^m P_{0,ij}^k(x)$ 
with $i,j$=$q,G$ and $k$=$S,NS$; and can be also found in Ref.~\cite{Altarelli:1977zs} (see Eqs.~(71-74)). The first row in the matrix of equation \eqref{eq:anomalous} leaves us with the standard one-loop DGLAP valence-quark evolution equation. The 2$\times$2 non-diagonal matrix block in \eqref{eq:anomalous} describes the evolution of the singlet components and makes also apparent how gluon and quarks become coupled. Indeed, one only needs to deal with the eigenvalue's problem for the matrix in Eq.~\eqref{eq:anomalous}, and its solutions can be formally written as
%----
\begin{eqnarray}
 \left( \begin{array}{cc}
1 & 0 \\ 
0 & {\bf P}^{-1} 
\end{array} \right) 
\left( \begin{array}{c}
\langle x_q^m \rangle_{\zeta}^{NS} \\ 
\langle x_q^m \rangle_{\zeta}^{S} \\
\langle x_G^m \rangle_{\zeta}^{S}
\end{array} \right) 
\ = \ 
\exp{\left( - \Gamma_D^{(m)} \int_{\ln{\zeta_0^2}}^{\ln{\zeta^2}} dt \; \frac{\alpha(e^t)}{4\pi}  
\right)}
 \left( \begin{array}{cc}
1 & 0 \\ 
0 & {\bf P}^{-1} 
\end{array} \right) 
\left( \begin{array}{c}
\langle x_q^m \rangle_{\zeta_0}^{NS} \\ 
\langle x_q^m \rangle_{\zeta_0}^{S} \\
\langle x_G^m \rangle_{\zeta_0}^{S}
\end{array} \right) 
 \ ,  
\label{eq:solution}
\end{eqnarray}
%----
where $\Gamma_D^{(m)}={\rm Diag}( \gamma_{0,qq}^{NS,(m)},\lambda_{+}^{(m)},\lambda_{-}^{(m)})$ is the matrix of eigenvalues for Eq.~\eqref{eq:anomalous} and ${\bf P}$ is the matrix which diagonalizes its 2$\times$2 non-diagonal block and, eventually, couples singlet quark and gluon distributions. At the leading order, $\alpha(\zeta)$ is taken from the integration of the 1-loop $\beta$-function and \eqref{eq:solution} can be thus displayed in terms of simple analytic expressions featuring the logaritmic running of the moments from $\zeta_0$ up to $\zeta$, both scales lying in the perturbative domain. However, our aim here (and so was in Ref.~\cite{Ding:2019lwe}) is evolving the valence-quark parton distribution obtained at a naturally nonperturbative hadronic scale, where the pion is only a bound sate of a dressed quark and a dressed antiquark, up to larger energy scales. To this goal, one can go beyond the leading-order approximation by recognising that an effective charge for the strong coupling in Eq.~\eqref{eq:master} can be defined such that the higher-order corrections become therein optimally neglected and, in order to make predictions, assuming then a phenomenological correspondence of this charge with a well-known effective coupling. 

\section{The interaction kernel and the process-independent effective strong coupling}

The interaction kernel used to get realistic solutions of the DSE gap equation for the quark propagator and for the BSA in Ref.~\cite{Ding:2019lwe}, and to compute thus the valence-quark parton distribution, is the one explained in Refs.~\cite{Qin:2011dd,Qin:2011xq}. This interaction has been found to coincide, in the infrared domain and within the error uncertainties, with a renormalisation-group-invariant (RGI) running-interaction resulting from contemporary studies of QCD's gauge sector\,\cite{Binosi:2014aea}, 
%-----
\begin{equation}
\label{allhatd}
\mathcal{I}(k^2) \ = \ k^2 \widehat{d}(k^2) \ = \
\frac{\alpha_{\rm T}(k^2)}{[1-L(k^2;\zeta^2)F(k^2;\zeta^2)]^2}\, ,
\end{equation}
%-----
with $\zeta$ still standing for the renormalisation scale;
$\widehat{d}(k^2)$ is a RGI function, owing to a sensible rearrangement of the diagramatic DSE expansion of the involved QCD Green's functions, within the approach given by the pinch technique and background field method (PTBFM)~\cite{Cornwall:1981zr,Binosi:2009qm}, as discussed in Ref.\,\cite{Aguilar:2009nf};
$F$ is the dressing function for the ghost propagator;
$L$ is a longitudinal piece of the gluon-ghost vacuum polarisation, playing a key role within the context of the PTBFM approach, and that vanishes at $k^2=0$\,\cite{Aguilar:2009nf}; and $\alpha_{\rm T}$ stands for the strong running coupling derived from the ghost-gluon vertex \cite{Sternbeck:2007br,Boucaud:2008gn,Aguilar:2009nf}, also called the ``Taylor coupling''  \cite{Blossier:2011tf,Blossier:2012ef,Blossier:2013ioa}. Further, on the ground of this running-interaction as a basic ingredient, the process-independent (PI) effective coupling, 
%---
\begin{equation}
\widehat{\alpha}_{\rm PI}(k^2)  = %\frac{\widehat{d}(k^2)} {\mathpzc{D}(k^2)} \ , 
\widehat{d}(k^2) / \mathpzc{D}(k^2)
\label{widehatalpha}
\end{equation}
%---
has been introduced and discussed in Refs.~\cite{Binosi:2016nme,Rodriguez-Quintero:2018wma,Binosietal}; where $1/{\mathpzc D}(k^2)$ is a mass-dimension-two RGI function defined from the gluon two-point Green's function, as explained therein. As can be seen in Fig.~\ref{fig:aPI}, this PI coupling is also found to describe well the world's data for the process-dependent Bjorken sum-rule effective charge (the roots of this striking coincidence, which opens a window for a direct experimental measure of the PI charge defined on the QCD's gauge sector, are largely discussed in Ref.~\cite{Binosi:2016nme}).  

Following Ref.\,\cite{Binosietal}, the coincidence of the PI effective coupling and the DSE interaction kernel, within the infrared domain, supports the assumption that the parametrisation
%----
\begin{equation}\label{eq:newalpha}
\frac{\alpha(\zeta^2)}{4\pi} = \left(\beta_0 \ln{\left(\frac{m_\alpha^2 + \zeta^2}{\Lambda^2_{\rm QCD}}\right)}\right)^{-1} \ , 
\end{equation}
%----
introduced in Ref.~\cite{Ding:2019lwe}, is the best candidate for the effective charge from Eq.~\eqref{eq:master}, 
where $\Lambda_{\rm QCD}$=234 MeV and $m_\alpha$=300 MeV are defined to make it coincides with the PI effective coupling, in the infrared, and smoothly connects with the pQCD tail of the kernel interaction in the ultraviolet. Then, as discussed in Ref.\,\cite{Ding:2019lwe}, $m_\alpha \gsim \Lambda_{\rm QCD}$ is a nonperturbative scale screening the soft gluon modes from interaction, which can be naturally identified with the hadronic scale $\zeta_H$. Therefore, plugging \eqref{eq:newalpha} into \eqref{eq:solution}, the valence-quark parton distribution can be unambigously evolved from $\zeta_H=m_\alpha$ up to $\zeta_5$, and then succesfully compared with the reanalysed E615 data\,~\cite{Wijesooriya:2005ir, Aicher:2010cb} (see Fig.~\ref{fig:PDFs}). Furthermore, at the hadronic scale, the pion is a two-valence-body bound-state with no explicit gluon or sea-quark contribution to the singlet distributions, only from valence dressed-quarks. Thus, QCD evolution 
effectively driven by \eqref{eq:newalpha} applied into \eqref{eq:solution} can be employed to estimate gluon and sea-quark distributions at any larger energy scale, accessible to experiment. The case for the $m=1$ Mellin moment (momentum fraction average) is particularly simple and the singlet distributions from \eqref{eq:solution} can be recast as ($N_f$=4)
%----
\begin{eqnarray}
\langle x_q \rangle^{S}_\zeta =     
\frac 3 7 + \frac 4 7 
\exp{\left( - \frac{56}{36\pi} \int_{\ln{\zeta_H^2}}^{\ln{\zeta^2}} dt \; \alpha(t)  \right)} 
% \nonumber \\ 
\, , \; \; \;   
\langle x_G \rangle^{S}_\zeta = \frac 4 7 
\left[ \rule[0cm]{0cm}{0.75cm} 
1 -  \exp{\left( - \frac{56}{36\pi} \int_{\ln{\zeta_H^2}}^{\ln{\zeta^2}} dt \; \alpha(t)  \right)} 
\right]  ; 
\label{eq:solMm1}
\end{eqnarray}
%---
which make apparent that, for $\Lambda_{\rm QCD}^2/\zeta^2 \to 0$,  sea-quark and gluon momentum fractions tend logarithmically to 3/4 and 4/7, respectively, while the valence-quark tends to 0\,\cite{Altarelli:1981ax}. 

\section{Conclusions}

We conclude by shortly sumarising. Continuum predictions for the pointwise behaviour of the pion's distribution functions for valence-quarks, gluons and sea are now consistently available for the first time, obtained within a Dyson-Schwinger-equations' approach and owing to the implementation of a symmetry-preserving interaction kernel. To this goal, we capitalised on a QCD's process-independent effective charge, driving the QCD evolution from a nonperturbative scale, unambigously defined by the freeze-out of interacting gluons, below its dynamical mass, up to any larger scale accessible to experiment. This leads to a parameter-free prediction of the pion's valence-quark distribution function that is in agreement with a modern analysis of the E615 data. The approach herein sketched can be potentially applied to extend the calculations to the spin-dependent structure functions and, beyond the kinematic forward limit, to the generalised parton distributions. 

\section*{Acknowledgements}

This discussion is based on work completed by an international collaboration involving many
remarkable people, to all of whom we are greatly indebted, and it is in connection with other
contributions in this volume, e.g. Craig D. Roberts'. J. R-Q would like to express his gratitude to 
the organisers of 27th International Nuclear Physics Conference (INPC 2019), who made possible my 
participation in a meeting that was both enjoyable and fruitful. The work has been partially 
supported by the Spanish ministry research project FPA2017-86380 and by the Jiangsu Province 
{\it Hundred Talents Plan for Professionals}.

\section*{References}
\bibliographystyle{iopart-num}
\bibliography{refs}

\providecommand{\newblock}{}
\begin{thebibliography}{10}
\expandafter\ifx\csname url\endcsname\relax
  \def\url#1{{\tt #1}}\fi
\expandafter\ifx\csname urlprefix\endcsname\relax\def\urlprefix{URL }\fi
\providecommand{\eprint}[2][]{\url{#2}}

\bibitem{Aguilar:2019teb}
Aguilar A~C {\em et~al.\/} 2019  (\textit{Preprint} \eprint{1907.08218})

\bibitem{Conway:1989fs}
Conway J~S {\em et~al.\/} 1989 {\em Phys. Rev.\/} {\bf D39} 92--122

\bibitem{Ezawa:1974wm}
Ezawa Z~F 1974 {\em Nuovo Cim.\/} {\bf A23} 271--290

\bibitem{Farrar:1975yb}
Farrar G~R and Jackson D~R 1975 {\em Phys. Rev. Lett.\/} {\bf 35} 1416

\bibitem{Berger:1979du}
Berger E~L and Brodsky S~J 1979 {\em Phys. Rev. Lett.\/} {\bf 42} 940--944

\bibitem{Hecht:2000xa}
Hecht M~B, Roberts C~D and Schmidt S~M 2001 {\em Phys. Rev.\/} {\bf C63} 025213
  (\textit{Preprint} \eprint{nucl-th/0008049})

\bibitem{Chang:2014lva}
Chang L, Mezrag C, Moutarde H, Roberts C~D, Rodríguez-Quintero J and Tandy P~C
  2014 {\em Phys. Lett.\/} {\bf B737} 23--29 (\textit{Preprint}
  \eprint{1406.5450})

\bibitem{Ding:2019lwe}
Ding M, Raya K, Binosi D, Chang L, Roberts C~D and Schmidt S~M 2019
  (\textit{Preprint} \eprint{1905.05208})

\bibitem{Wijesooriya:2005ir}
Wijesooriya K, Reimer P~E and Holt R~J 2005 {\em Phys. Rev.\/} {\bf C72} 065203
  (\textit{Preprint} \eprint{nucl-ex/0509012})

\bibitem{Aicher:2010cb}
Aicher M, Schafer A and Vogelsang W 2010 {\em Phys. Rev. Lett.\/} {\bf 105}
  252003 (\textit{Preprint} \eprint{1009.2481})

\bibitem{Bednar:2018mtf}
Bednar K~D, Cloët I~C and Tandy P~C 2018  (\textit{Preprint}
  \eprint{1811.12310})

\bibitem{Barry:2018ort}
Barry P~C, Sato N, Melnitchouk W and Ji C~R 2018 {\em Phys. Rev. Lett.\/} {\bf
  121} 152001 (\textit{Preprint} \eprint{1804.01965})

\bibitem{Mezrag:2014jka}
Mezrag C, Chang L, Moutarde H, Roberts C~D, Rodríguez-Quintero J, Sabatié F and
  Schmidt S~M 2015 {\em Phys. Lett.\/} {\bf B741} 190--196 (\textit{Preprint}
  \eprint{1411.6634})

\bibitem{Binosi:2016nme}
Binosi D, Mezrag C, Papavassiliou J, Roberts C~D and Rodriguez-Quintero J 2017
  {\em Phys. Rev.\/} {\bf D96} 054026 (\textit{Preprint} \eprint{1612.04835})

\bibitem{Rodriguez-Quintero:2018wma}
Rodríguez-Quintero J, Binosi D, Mezrag C, Papavassiliou J and Roberts C~D 2018
  {\em Few Body Syst.\/} {\bf 59} 121 (\textit{Preprint} \eprint{1801.10164})

\bibitem{Gao:2017mmp}
Gao F, Chang L, Liu Y~X, Roberts C~D and Tandy P~C 2017 {\em Phys. Rev.\/} {\bf
  D96} 034024 (\textit{Preprint} \eprint{1703.04875})

\bibitem{Ding:2018xwy}
Ding M, Raya K, Bashir A, Binosi D, Chang L, Chen M and Roberts C~D 2019 {\em
  Phys. Rev.\/} {\bf D99} 014014 (\textit{Preprint} \eprint{1810.12313})

\bibitem{Burkardt:2002uc}
Burkardt M, Ji X~d and Yuan F 2002 {\em Phys. Lett.\/} {\bf B545} 345--351
  (\textit{Preprint} \eprint{hep-ph/0205272})

\bibitem{Diehl:2003ny}
Diehl M 2003 {\em Phys. Rept.\/} {\bf 388} 41--277 (\textit{Preprint}
  \eprint{hep-ph/0307382})

\bibitem{Xu:2018eii}
Xu S~S, Chang L, Roberts C~D and Zong H~S 2018 {\em Phys. Rev.\/} {\bf D97}
  094014 (\textit{Preprint} \eprint{1802.09552})

\bibitem{Rayaetal}
Raya K and {\it et al} {\em {\it in preparation.}\/}

\bibitem{Altarelli:1977zs}
Altarelli G and Parisi G 1977 {\em Nucl. Phys.\/} {\bf B126} 298--318

\bibitem{Qin:2011dd}
Qin S~x, Chang L, Liu Y~x, Roberts C~D and Wilson D~J 2011 {\em Phys.Rev.\/}
  {\bf C84} 042202 (\textit{Preprint} \eprint{1108.0603})

\bibitem{Qin:2011xq}
Qin S~x, Chang L, Liu Y~x, Roberts C~D and Wilson D~J 2012 {\em Phys. Rev.\/}
  {\bf C85} 035202 (\textit{Preprint} \eprint{1109.3459})

\bibitem{Binosi:2014aea}
Binosi D, Chang L, Papavassiliou J and Roberts C~D 2015 {\em Phys. Lett.\/}
  {\bf B742} 183--188 (\textit{Preprint} \eprint{1412.4782})

\bibitem{Cornwall:1981zr}
Cornwall J~M 1982 {\em Phys.Rev.\/} {\bf D26} 1453

\bibitem{Binosi:2009qm}
Binosi D and Papavassiliou J 2009 {\em Phys.Rept.\/} {\bf 479} 1--152 245
  pages, 92 figures (\textit{Preprint} \eprint{0909.2536})

\bibitem{Aguilar:2009nf}
Aguilar A~C, Binosi D, Papavassiliou J and Rodriguez-Quintero J 2009 {\em Phys.
  Rev.\/} {\bf D80} 085018 (\textit{Preprint} \eprint{0906.2633})

\bibitem{Sternbeck:2007br}
Sternbeck A {\em et~al.\/} 2007 {\em PoS\/} {\bf LAT2007} 256
  (\textit{Preprint} \eprint{0710.2965})

\bibitem{Boucaud:2008gn}
Boucaud P, De~Soto F, Leroy J, Le~Yaouanc A, Micheli J, P\`ene O and
  Rodr\'{\i}guez-Quintero J 2009 {\em Phys.Rev.\/} {\bf D79} 014508
  (\textit{Preprint} \eprint{0811.2059})

\bibitem{Blossier:2011tf}
Blossier B, Boucaud P, Brinet M, De~Soto F, Du X {\em et~al.\/} 2012 {\em
  Phys.Rev.\/} {\bf D85} 034503 (\textit{Preprint} \eprint{1110.5829})

\bibitem{Blossier:2012ef}
Blossier B, Boucaud P, Brinet M, De~Soto F, Du X, Morenas V, Pene O, Petrov K
  and Rodriguez-Quintero J 2012 {\em Phys. Rev. Lett.\/} {\bf 108} 262002
  (\textit{Preprint} \eprint{1201.5770})

\bibitem{Blossier:2013ioa}
Blossier B {\em et~al.\/} (ETM Collaboration) 2014 {\em Phys.Rev.\/} {\bf D89}
  014507 (\textit{Preprint} \eprint{1310.3763})

\bibitem{Binosietal}
Binosi D and {\it et al} {\em {\it in preparation.}\/}

\bibitem{Chouika:2017rzs}
Chouika N, Mezrag C, Moutarde H and Rodríguez-Quintero J 2018 {\em Phys.
  Lett.\/} {\bf B780} 287--293 (\textit{Preprint} \eprint{1711.11548})

\bibitem{Deur:2016tte}
Deur A, Brodsky S~J and de~Teramond G~F 2016 {\em Prog. Part. Nucl. Phys.\/}
  {\bf 90} 1--74 (\textit{Preprint} \eprint{1604.08082})

\bibitem{Altarelli:1981ax}
Altarelli G 1982 {\em Phys. Rept.\/} {\bf 81} 1

\end{thebibliography}

\end{document}